# Ultra-broadband acoustic absorber based on periodic acoustic rigid-metaporous composite array

Dongguo Zhang[1, #], Fei Sun[1, #, *], and Yichao Liu[1]

*1 Key Lab of Advanced Transducers and Intelligent Control System, Ministry of Education and Shanxi Province, College of Electronic Information and Optical Engineering, Taiyuan University of Technology, Taiyuan, 030024 China*

\# Dongguo Zhang and Fei Sun contributed equally to this work.

\* Corresponding author: sunfei@tyut.edu.cn

**Abstract**

**To address the increasingly serious issue of noise pollution, we propose an ultra-broadband and wide-angle acoustic absorber based on a periodic acoustic rigid-metaporous composite array. Numerical simulation results verify the broadband good acoustic absorption performance of the proposed absorber, which can achieve an average absorption coefficient of approximately 90.9% within the frequency band from 500 to 4000 Hz with incident angles ranging from -75 to +75 degrees, thus compensating for the shortcomings of traditional acoustic absorbers that are not as effective at low frequencies. This work will provide a new approach for ultra-broadband and wide-angle acoustic wave absorption and noise suppression.**

## 1. Introduction

Noise pollution is a pervasive issue that significantly affects the well-being of communities. It is essential to address this challenge as it can lead to various health problems, including hearing loss, sleep disorders, and psychological stress. Therefore, it is necessary to find effective means for broadband noise suppression. Traditional methods for noise suppression have primarily focused on the use of barriers and absorbent materials. Acoustic barriers, such as walls and fences, are commonly employed to block the propagation of noise. These barriers, however, often have limited effectiveness due to their inability to absorb a wide range of frequencies, leading to a narrow bandwidth of noise reduction [1]. Another traditional approach involves the use of absorbent materials, which work by dissipating the energy of acoustic waves. These materials, such as porous foams and fibrous materials, can be effective at certain frequencies but struggle to provide broadband absorption [2],[3]. The challenge lies in the fact that these materials are often designed to absorb specific frequencies, and their effectiveness diminishes outside of these narrow bandwidths.

In recent years, acoustic metamaterials/metasurfaces [4]-[11], which can achieve acoustic wave manipulation effects unattainable by natural materials [12]-[16], have introduced new methods for acoustic control and have shown promise in addressing the limitations of traditional noise suppression methods [17]. These advanced materials are engineered to exhibit acoustic properties not found in nature, thereby allowing for greater control over the absorption and transmission of acoustic wave. Currently, various acoustic metamaterials/metasurfaces have been applied to attenuate noise. The fundamental approach to enhance the acoustic absorption bandwidth of a composite structure using acoustic metamaterials/metasurfaces is to combine or nest metamaterial units with different acoustic absorption frequencies. This can be achieved by integrating multiple metamaterials/metasurfaces with distinct inherent frequencies, such as space-coiled metamaterials [18],[19] or Helmholtz resonators [20]-[23]. By doing so, the entire composite structure can absorb a wider range of acoustic frequencies, thereby achieving a relatively broad bandwidth of acoustic absorption. However, the working bandwidth of these composite structures based on acoustic metamaterials/metasurfaces is still limited, whose absorption coefficients are generally around 80% within the sensitive frequency range of human hearing, which spans from 500 to 4000 Hz [24].

To further improve the acoustic absorption frequency range and efficiency, porous materials can be integrated with acoustic metamaterials/metasurfaces. Combining porous materials with acoustic metamaterials/metasurfaces results in the formation of metaporous materials [25]-[32], where acoustic energy is trapped and dissipated inside the composite acoustic structure while being absorbed by porous material. This allows for the utilization of both the acoustic absorption response of the acoustic metamaterials/metasurfaces resonance and the inherent acoustic absorption response of the porous materials, thereby achieving a wider acoustic absorption frequency range. For example, porous materials can be combined with Helmholtz resonators to obtain an improved acoustic absorption bandwidth, possessing both the absorption response of the Helmholtz resonators and the inherent acoustic absorption response of the porous materials [25]-[27]. Combining porous materials with periodic arrangements of rigid partitions forms metaporous layers, allowing resonance modes in the layer to manifest at various frequencies [28]-[32]. However, the bandwidth of these acoustic absorbers combined with porous foam still cannot cover the entire sensitive frequency range from 500 to 4000 Hz. Specifically, most of their operation bandwidth is not lower than 1000 Hz [25],[28]-[30]. While some structures introduce additional absorption peaks around several hundred hertz [27],[31],[32] to expand the bandwidth, none have achieved an absorption coefficient exceeding 80% across the range from 500 to 4000 Hz.

To address this issue, this study introduces a novel periodic acoustic rigid-metaporous composite array, as depicted in Fig. 1(a), aiming to develop an ultra-broadband acoustic absorber that exhibits high acoustic absorption efficiency. Its excellent acoustic absorption performance has been verified through numerical simulations, showing an average absorption coefficient of approximately 90.9% across a frequency range from 500 to 4000 Hz with incident angles ranging from -75 to 75 degrees.

## 2. Design and numerical simulations

As shown in Fig. 1(a), the proposed acoustic absorber is periodic acoustic rigid-metaporous composite array, with a detailed view of a single unit presented in the enlarged inset. The individual units within the array structure of the acoustic absorber are formed by a resin (depicted in gray, serving as hard wall boundary condition) that encases an acoustic tunnel. Inside this tunnel, a wedge-shaped segment of porous material is filled, which is illustrated in blue and typically made from melamine foam. Each unit within the array mirrors this configuration, consisting of a rigid acoustic tunnel that houses the wedge-shaped porous insert. Additionally, each tunnel incorporates sub-wavelength protrusion structures to modulate the total acoustic path and the refractive index [34]. These protrusions are also arranged periodically, as detailed in secondary magnified illustration of Fig. 1(a), where the protrusion length by $p$, the spacing between protrusions by $d$, and the uniform thickness of both the protrusions and the tunnel by $t$. The tunnel's cross-sectional width $w$ is designed to be much smaller than the wavelength, to ensure that there is only one mode in the acoustic tunnel [34]-[37]. The height $h$ of each unit is consistent across the entire absorber, ensuring uniformity in design and function.

Each subwavelength acoustic tunnel with protrusion structures filled by wedge-shaped porous materials (e.g., melamine foam) exhibits significant capability in impedance matching with the surrounding air over a wide frequency band. This ensures that when acoustic waves incident onto the periodic acoustic rigid-metaporous composite array shown in Fig. 1(a), there are no reflected waves across the broad frequency range. Additionally, the geometric parameters of protrusion structures, particularly the ratio of the protrusion length to the tunnel's cross-sectional width (i.e., $p/w$), play a key role in modulating the total acoustic path and the refractive index of the tunnel [34].

As an example, a frequency of $f = 1000$ Hz is chosen to demonstrate the acoustic path modulating function of protrusion structures, as depicted in Fig. 1(b). The simulated results show that, with other parameters fixed, as the ratio of the protrusion length to the tunnel's cross-sectional width $p/w$ increases, the wavelength of acoustic waves propagating within the tunnel shortens meaning that the acoustic path increases for the same geometric length. By calculating the wavelength $\lambda$ of acoustic waves inside the tunnel and the ratio of $\lambda_0$ (acoustic wavelength when propagating in air) to $\lambda$, the equivalent refractive index $n_r$ of the tunnel can be derived, which is a function of the ratio $p/w$. As shown in Fig. 1(c), for a fixed $p/w$, the refractive index at different frequencies is almost the same. This implies that the equivalent refractive index of the acoustic sub-wavelength tunnel with protrusion structures is nearly dispersionless (i.e., it exhibits characteristics of a broad bandwidth operation), which can be tuned by changing the geometric parameters of the protrusion structures within the tunnels, specifically the ratio $p/w$.

As the equivalent refractive index increases, the acoustic path of the acoustic wave propagating in the tunnel increases, and the attenuation of the acoustic wave in the tunnel with melamine foam also increases accordingly. To quantitatively study the acoustic absorption performance of the designed acoustic absorber, acoustic absorption coefficient is introduced, which can be defined as $\alpha = 1 - R^2$. Here, $R = p_r/p_i$ is the reflection coefficient. $p_r$ and $p_i$ are the averaged reflected acoustic pressure and incident acoustic pressure, respectively.

To show the absorption property varies with the ratio $p/w$, we simulated the effect of the absorption coefficient changing with $p/w$ during normal incidence of acoustic waves. The geometric parameters for the acoustic absorber in Fig. 1(a) are designed as follows: $h = 81$ mm, $t = 1$ mm, $d = 5$ mm, $w = 9$ mm, $p$ varies from 0 to 3 mm. Fig. 1(d) shows the absorption coefficient for the proposed absorber array varying with the ratio $p/w$ in the frequency range of 500 to 4000 Hz for acoustic waves at normal incidence. In most cases, the absorption coefficient increases with the increase of the ratio $p/w$, which confirms the previous inference that the absorption coefficient increases with the refractive index. However, in some rare cases, due to relatively strong reflections in certain frequency ranges (such as around 3200 Hz), the absorption coefficient slightly decreases at high $p/w$ ratios. Based on this analysis, we recommend setting $p/w$ between 0.30 and 0.35 as the most appropriate range.

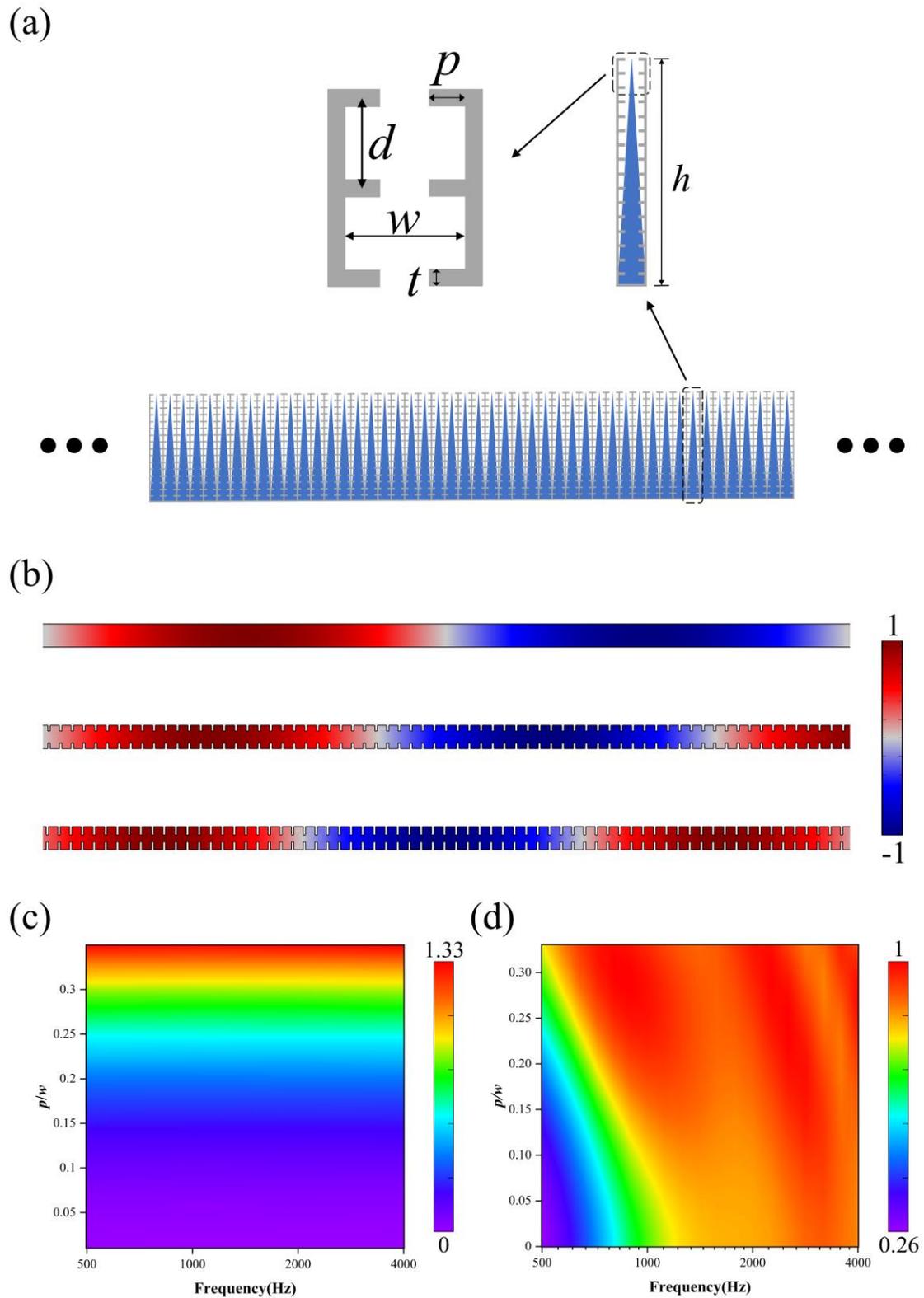

**Fig. 1.** (a) Schematic diagram of the acoustic absorber. The gray portion represents the solid structure, and the blue portion represents porous foam. (b) Acoustic pressure field distributions at 1000 Hz that vary with the ratio of $p/w$. From top to bottom, the values of $p/w$ are 0, 0.24, and 0.35, corresponding to refractive index $n_r$ = 1, 1.147, 1.271. (c) Refractive index $n_r$ as a function of $p/w$ from 500 to 4000 Hz. (d) The absorption coefficient varying with $p/w$ in the frequency range of 500 to 4000 Hz for acoustic waves at normal incidence.

Numerical simulation set up is shown in Fig. 2(a). The simulation area is surrounded by perfectly matched layer to avoid boundary reflection and is used to simulate an infinite free space. The incident wave is an acoustic Gaussian beam with a range of frequencies (from 500 to 4000 Hz). The porous materials are chosen as melamine foam, whose equivalent bulk modulus and mass density are described by the Johnson–Champoux–Allard model [38]-[40], where porosity $\varepsilon_p$ = 0.995, flow resistivity $R_f$ = $10.5 \times 10^3$ Pa·s/m$^2$, viscous characteristic length $L_v$ = 240 um, thermal characteristic length $L_{th}$ = 470 μm and tortuosity factor $\tau_\infty$ = 1.0059. Acoustic rigid tunnels, composed of resin, are modeled as hard wall boundary conditions in all simulations. The background area is filled up with air, whose equivalent bulk modulus and mass density are $1.01 \times 10^5$ Pa and 1.29 kg/m$^3$, respectively.

All the numerical simulations performed in this study are conducted by COMSOL Multiphysics 5.6. All simulations are 2D cases where the poroacoustics module with steady-state solver is used to calculate the acoustic pressure distributions. Free triangle meshing is used, where the maximum grid is one-tenth of the minimum working wavelength ($\lambda_{min}$ = 85.75 mm) to ensure accuracy of simulations.

$p$ = 2.98 mm is chosen as an example to illustrate the broadband acoustic absorption performance of the proposed periodic acoustic rigid-metaporous composite array. Other parameters are the same as the above settings. The total normalized acoustic pressure distribution for different frequencies is numerically simulated, as shown in Fig. 2(b)-(e). The simulation results preliminarily indicate that for acoustic Gaussian beams with frequencies of 500, 1000, 2000, and 4000 Hz incident at 30 degrees onto the proposed acoustic absorber, the reflection wave almost disappears, indicating that most of the incident acoustic waves are absorbed by the absorber, thereby achieving broadband acoustic absorption. It should be noted that the faint interference fringes appearing in Fig. 2(b)-(e) are caused by the interference between the incident wave and the weak reflection wave, indicating that the absorption efficiency of the structure is not 100%.

To verify the wide-angle acoustic absorption effect of the acoustic absorber, a frequency of $f$ = 1000 Hz is chosen as a demonstrating example. Fig. 3 shows the simulated total normalized acoustic pressure distribution as acoustic Gaussian beams vary in angle of incidence from 0 to 75 degrees on the designed acoustic absorber. Within the range of 0 to 60 degrees, the reflection wave almost disappears, which indicate excellent acoustic absorbing performance. Even at an incidence angle of 75 degrees, the intensity of the reflected wave is significantly reduced compared to the incident wave, demonstrating the wide-angle acoustic absorption performance of the acoustic absorber. Due to the symmetry of the proposed acoustic absorber along the *y*-axis, the simulation results for acoustic waves incident from 0 to -75 degrees are completely symmetrical to those from 0 to 75 degrees. Therefore, Fig. 3 only presents the simulation results from 0 to 75 degrees.

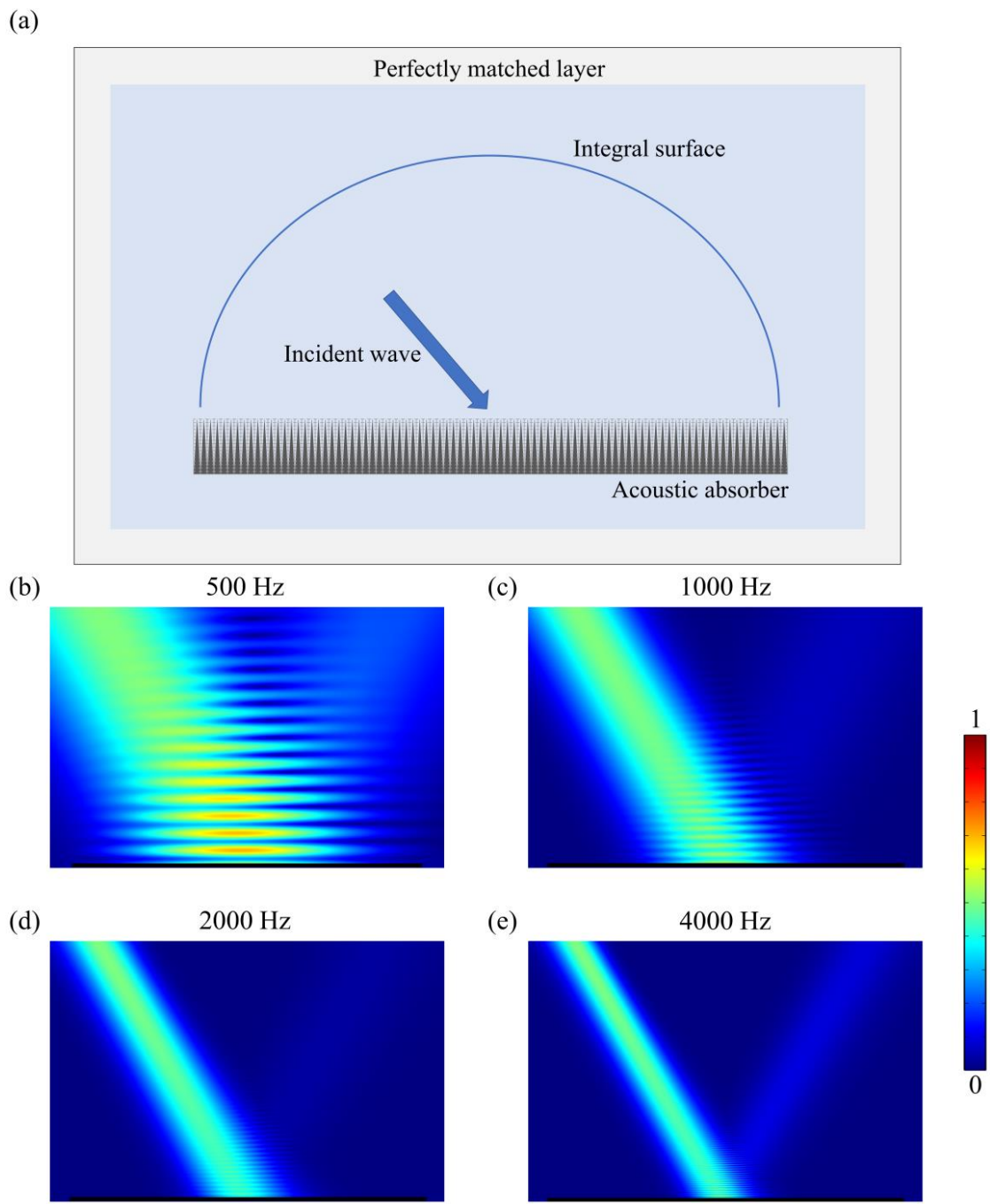

**Fig. 2.** (a) Simulation domain setup. (b)-(e) The normalized acoustic pressure when a Gaussian acoustic beam incidents at 30 degrees onto the acoustic absorber at 500, 1000, 2000, 4000 Hz. In these simulations, geometric parameters for the acoustic absorber are $h$ = 81 mm, $t$ = 1 mm, $d$ = 5 mm, $w$ = 9 mm, $p$ = 2.98 mm.

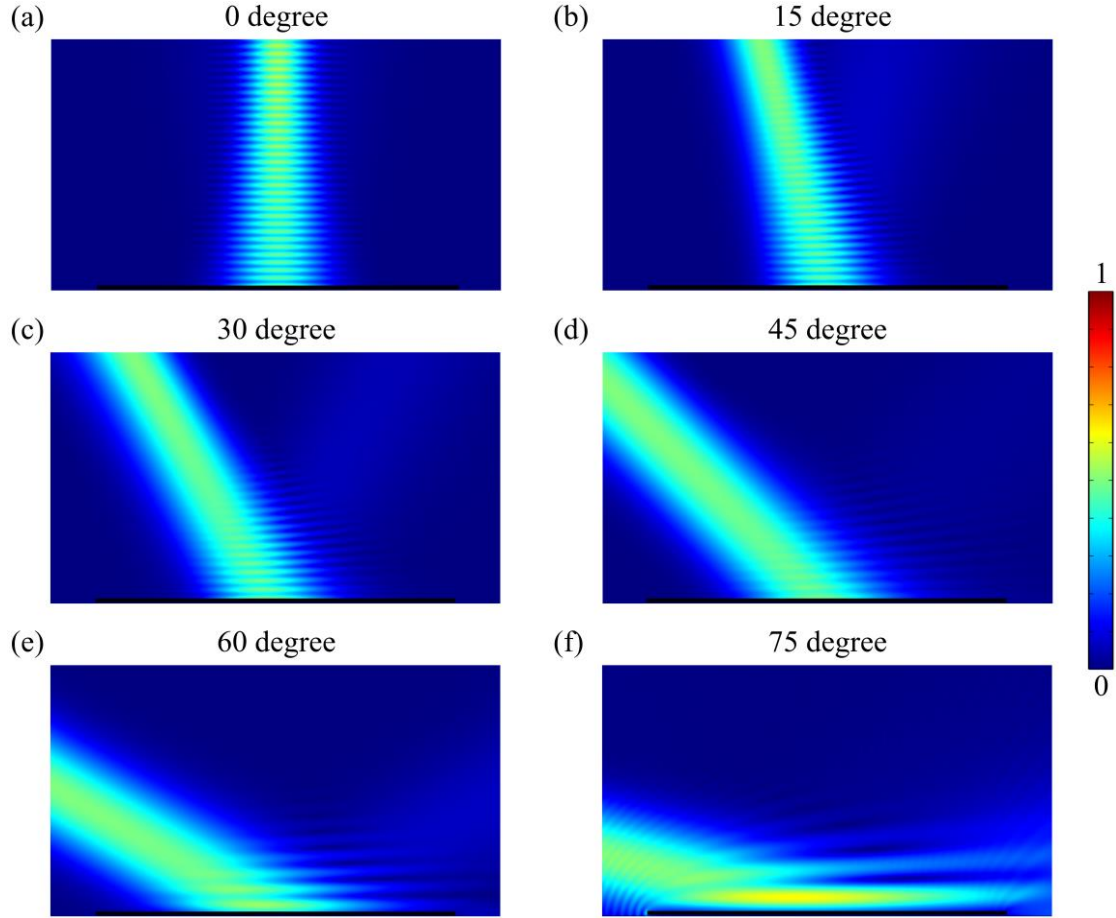

**Fig. 3.** The normalized acoustic pressure when a Gaussian acoustic beam with a frequency of 1000 Hz incidents onto the acoustic absorber by 0, 15, 30, 45, 60 and 75 degrees. Other parameters in these simulations are the same as these used in Fig. 2(b)-(e).

## 3. Results

The simulated absorption coefficient of an acoustic Gaussian beam ranging from 500 to 4000 Hz, incident from -75 to 75 degrees onto the periodic acoustic rigid-metaporous composite array. As the proposed acoustic absorber exhibits symmetry along the *y*-axis, the simulation results for acoustic waves incident from 0 to -75 degrees are completely symmetrical to those from 0 to 75 degrees (with corresponding equal acoustic absorption coefficient). Therefore, Fig. 4(a) only presents the simulated acoustic absorption coefficient from 0 to 75 degrees. It can be observed that in most cases, the absorption coefficient exceeds 90%, and even at low frequencies around 500 Hz, it still can reach 80%. Compared to the acoustic absorption effect of wedge-shaped porous foam array without periodic acoustic rigid tunnels shown in Fig. 4(b), the absorption coefficient has significantly improved across all frequency bands, and the effective incident angle has noticeably increased. The simulated results in Fig. 4(a) show that the periodic acoustic rigid-metaporous composite array can provide an average absorption coefficient of approximately 90.9% across a frequency range from 500 to 4000 Hz with incident angles ranging from 0 to 75 degrees. Therefore, the proposed acoustic absorber can achieve excellent acoustic absorption effects in environments where broadband noise is present.

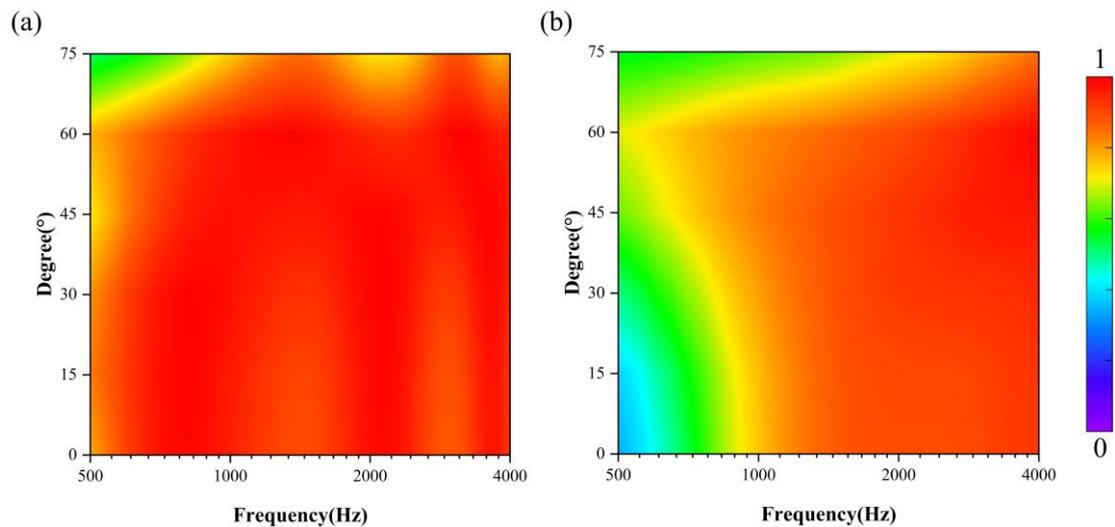

**Fig. 4.** Simulation results of the acoustic absorption coefficient for acoustic waves ranging from 500 to 4000 Hz incident at angles from 0 to 75 degrees onto (a) the periodic acoustic rigid-metaporous composite array and (b) wedge-shaped melamine foam array without periodic acoustic rigid tunnels. Other parameters in this simulation are the same as these used in Fig. 2.

## 4. Conclusions

In conclusion, the periodic acoustic rigid-metaporous composite array is designed to achieve an ultra-broadband and wide-angle acoustic absorbing, which can match the impedance of air well to reduce reflectivity and absorb acoustic waves through the porous foam in the tunnels, showing great potential for broad applications in noise attenuation. The proposed acoustic absorber boasts an exceptionally high bandwidth, with the capability to attain an average absorption coefficient of approximately 90.9% for acoustic waves within a frequency band ranging from 500 to 4000 Hz, across incident angles ranging from -75 to 75 degrees. The findings of this work offer a novel approach to ultra-broadband and wide-angle acoustic wave absorption, providing a new strategy for the noise suppression. This development holds promise for the creation of more effective noise control solutions, contributing to a quieter and more comfortable environment.

**Funding**

This work is supported by the National Natural Science Foundation of China (Nos. 12274317, 12374277, 61971300, 61905208, and 11604292), Basic Research Project of Shanxi Province 202303021211054, and University Outstanding Youth Foundation of Taiyuan University of Technology.

**CRediT authorship contribution statement**

**Dongguo Zhang:** Conceptualization, Formal analysis, Writing - original draft, Validation. **Fei Sun:** Conceptualization, Methodology, Writing - review & editing, Formal analysis, Supervision, Funding acquisition. **Yichao Liu:** Methodology, Funding acquisition.

**Declaration of competing interest**

The authors declare that they have no known competing financial interests or personal relationships that could have appeared to influence the work reported in this paper.